\documentclass[aps,twocolumn]{revtex4}
\usepackage{amsmath}
\usepackage{graphicx}

\newcommand{\eq}{\begin{equation}}
\newcommand{\fine}{\end{equation}}
\begin{document}
\title{Entanglement, EPR correlations and mesoscopic quantum superposition by the
high-gain quantum injected parametric amplification }
\author{Marco Caminati$^{1}$,\ Francesco De Martini$^{1}$, Riccardo Perris$^{1}$,
Fabio Sciarrino$^{2,1}$, Veronica Secondi$^{1}$}
\address{$^{1}$Dipartimento di Fisica and Consorzio Nazionale Interuniversitario per
le Scienze Fisiche della \ Materia, Universit\'{a} ''La Sapienza'',
Roma
00185, Italy\\
$^{2}$Centro di\ Studi e Ricerche ''Enrico Fermi'', Via Panisperna
89/A, Compendio del Viminale, Roma 00184, Italy}

\begin{abstract}
We investigate the multiparticle quantum superposition and the persistence
of multipartite entanglement of the quantum superposition generated by the
quantum injected high-gain optical parametric amplification of a single
photon. The physical configuration based on the optimal universal quantum
cloning has been adopted to investigate how the entanglement and the quantum
coherence of the system persists for large values of the nonlinear
parametric gain $g$.
\end{abstract}

\maketitle

\section{Introduction}

In recent years, a large number of experiments aimed at the verification of
fundamental aspects of quantum mechanics, such as quantum nonlocality, have
been realized by adopting photon particles mutually interacting through
non-linear optical (NLO) process. In addition sophisticated NLO methods have
been extended to relevant investigations and realizations in the domain of
the emerging new sciences of Quantum Information (QI) and Quantum
Communication. In particular one of such methods, the quantum injected
nonlinear (NL)\ parametric amplification (QIOPA)\ of single photon states,
was particularly fruitful since it was adopted to provide the first
experimental realization of the quantum cloning transformation, a
fundamental QI\ concept \cite{Scar05,Cerf05,DeMa05}. This one provides the
optimal distribution of the information contained in a quantum system, e.g.
a $N$ quantum-bit state, or {\it qubit} onto a system of higher dimension, $%
M>N$ \cite{DeMa98,DeMa98b}. By virtue of the isomorphism existing
between any logic qubit associated with spin-$1/2$ and the
polarization state of a single photon, there it is generally
supposed that $N$ photons, identically
prepared in an arbitrary state of polarization ($\left| \phi \right\rangle $%
), are injected into the amplifier on the input mode ${\bf k}_{1}$ . The
amplifier then generates on the output cloning mode (C) $M>N$ copies, or
{\it clones} of the input qubit $\left| \phi \right\rangle $. Moreover, in
the case of mode-nondegenerate QIOPA, the device simultaneously generates $%
M-N$ states $\left| \phi ^{\perp }\right\rangle $ on the output {\it %
anticloning} mode ${\bf k}_{2}$ (AC) thus realizing an universal quantum NOT
gate \cite{DeMa02}.

In the last years, the QIOPA scheme has been at the basis of experimental
realizations of the $1\rightarrow 2$ universal optimal quantum cloning
machine (UOQCM) \cite{DeMa02,Lama02,Pell03,DeMa04,DeMa05} and of the $%
1\rightarrow 3$ phase covariant quantum cloning machine (PQCM) \cite{Scia05}%
. These tests, carried out in low power linearized conditions, i.e.
with very low values of the NL parametric gain parameter $g<<1$,
were followed recently by a series of \ OPA\ works, carried out in
absence and in presence of quantum injection, which realized the
high gain (HG) spontaneous and stimulated generation of large number
of output photons $M$ \cite {Eise04,DeMa05b,Cami06}. Within this new
$N\rightarrow M$ cloning endeavor, a multi-photon superposition
entangled state was generated, indeed a Schroedinger Cat state
\cite{Schleich01,Schroe35}. By virtue of the information-preserving
(i.e. coherence-preserving) property of the parametric process, this
implied the deterministic transferal of the well accessible and
easily achievable quantum superposition condition affecting any
input single-particle qubit to a ''mesoscopic'', i.e.
multi-particle, amplified quantum-state \cite{DeMa05b}.

In the present article, we investigate theoretically the nature of
quantum injected optical parametric amplification in high gain
regime. In particular we intend to investigate the most important
property
of the process, i.e. the entanglement of the output modes in the {\it %
multiparticle} condition. We show how this difficult task can be undertaken
by application of a novel technique, here referred to as ''{\it pair} {\it %
extraction technique}'', adopted to investigate the multiphoton states
generated by high-gain spontaneous parametric down-conversion (SPDC)

\begin{figure}[t]
\includegraphics[scale=.4]{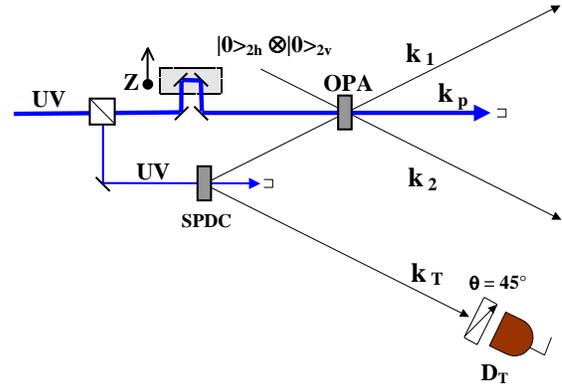} \caption{(Color on line) Schematic diagram of the single photon {\it %
quantum injected} optical parametric amplifier (QIOPA). The
injection is provided by an external spontaneous parametric down
conversion source of polarization entangled photon states
\cite{DeMa98}.The losses are simulated
by the insertion of a beamsplitter $(BS)$ over each propagation mode ${\bf k}%
_{i}$. Inset:\ each beamsplitter couples input modes
$a_{IN}^{\dagger }$, with unpopulated ones $(a_{IN}^{\prime \dagger
})$.} \label{fig1}
\end{figure}

In case of a bipartite entanglement, e.g. established over the
output cloning and anticloning modes ${\bf k}_{1}$ and ${\bf
k}_{2}$, the basic idea is to investigate a two-photon field
component ''extracted'' out of \ the output multi-photon field and
then infer the entanglement properties of the original field.
Conceptually, the basic argument underlying this method consists of
the impossibility of creating or enhancing the entanglement by any
local operation, e.g. in this case by induced losses \cite{Vedr98}.
In order to do that, in our case the adopted theoretical model takes
into account\ the propagation losses leading to the imperfect
detection of the output field. In the SPDC case the explicit form
for the two-photon output state has been found to exhibit a Werner
state structure, i.e., consisting of a\ weighted mixture of a
maximally entangled (singlet) state with a fully depolarized state
\cite{Wern89}. This structure is resilient to losses for any value
of the nonlinear gain parameter $g$. A similar approach will be
applied to investigate theoretically and experimentally the
bipartite entanglement in the high-gain QIOPA process. The
theoretical model enables to obtain the explicit form of the
two-photon output density matrix for any value of $g$. More
precisely, the application of the method goes as follows. Let's
start from a polarization entangled pair of photons associated with
the output modes ${\bf k}_{1}$ and ${\bf k}_{T}$: Figure1. The
photon created over ${\bf k}_{T}$ freely propagates and is detected
by the phototube $D_{T}$ while the photon associated to mode ${\bf
k}_{1}$ is injected into the optical parametric amplifier. As a
result of the coherence-preserving nature of the amplification
process the quantum correlations between the initial two photon
entangled states are transferred into correlations between the
photon on mode ${\bf k}_{T}$ (hereafter referred to as \ the
''trigger'') and the output field of the QIOPA\ device.

In order to characterize the field generated over the three output modes $%
{\bf k}_{T}$, ${\bf k}_{1}$ and ${\bf k}_{2}$ in high gain regime,
we explicitly derive the reduced density matrix of the three photon
states obtained over all the output modes adopting the {\it \ pair}
{\it extraction technique}. Hence we shall investigate the
characteristics of each bipartite systems: trigger-cloning,
trigger-anticloning and cloning-anticloning modes. Each subsystem
has been found to exhibit a Werner structure. By adopting this model
we are able to analyze the general entanglement properties of the
overall 3 mode system. Furthermore, this leads to the demonstration
of the persistence of entanglement for the two systems
trigger-cloning and cloning-anticloning for any large value of the
gain parameter $g$, in our opinion the main result of the present
work \cite {Vedr98}.

The article is structured as follows. In Section II we present the method to
extract from multiphoton states generated by amplification process two
photons components making use of lossy channel and imperfect detection. In
Section III we analyze the quantum correlations between trigger, cloning and
anticloning, realizing a model for the three qubits density matrix with the
same projection scheme. Finally, Section IV is devoted to the discussions of
agreement with the experimental apparatus and the perspectives of the
present research.

\section{QIOPA in the pair-extraction regime}

In this Section, we analyze the amplified multiphoton quantum injected state
with particular emphasis on the effect of losses on bipartite correlations.
Let us consider the experimental scheme reported in Fig.1. A single photon
(mode ${\bf k}_{1})$ is injected into the non-linear (NL) crystal, typically
a BBO ($\beta $-barium-borate), cut for Type II phase matching and excited
by a sequence of UV (ultra-violet)\ mode-locked laser pulses of wavelength
(wl) $\lambda _{p}$ propagating along the mode ${\bf k}_{P}$ . The relevant
modes of the NL 3-wave interaction driven by the UV pulses associated with
mode ${\bf k}_{p}$ are the two spatial modes with wave-vector (wv) ${\bf k}%
_{i}$, $i=1,2$, each one supporting the two horizontal $(H)$ and vertical $%
(V)$ polarizations of the interacting photons. The QIOPA is $\lambda $%
-degenerate, i.e. the interacting photons have the same wl's $\lambda
=2\lambda _{p}.$ The injected single photon is provided by an external
spontaneous parametric down conversion source of biphoton states \cite
{DeMa98}.

The Hamiltonian of the parametric down-conversion process in the interaction
picture reads \cite{DeMa98,DeMa05}
\begin{equation}
\hat{H}=i\kappa (\hat{a}_{1H}^{\dagger }\hat{a}_{2V}^{\dagger }-\hat{a}%
_{1V}^{\dagger }\hat{a}_{2H}^{\dagger })+h.c.  \label{hamiHV}
\end{equation}
where $\hat{a}_{ij}^{\dagger }$ represents the creation operators associated
to the spatial propagation mode ${\bf k}_{i}$, with polarization $j=\{H,V\}$%
. The NL\ coupling constant $\kappa $ depends on the nonlinearity of the
crystal and is proportional to the amplitude of the pump beam. It has been
theoretically shown \cite{DeMa00,Simo00},\ that $\widehat{H}$ is invariant
under simultaneous general $SU(2)$ transformations of the polarization
vectors for modes ${\bf k}_{1}$ and ${\bf k}_{2}.$ We may, then, cast the
above expression (\ref{hamiHV}) in the following form:
\begin{equation}
\hat{H}=i\kappa \left( \widehat{a}_{\Psi }^{\dagger }\widehat{b}_{\Psi \perp
}^{\dagger }-\widehat{a}_{\Psi \perp }^{\dagger }\widehat{b}_{\Psi
}^{\dagger }\right) +h.c.  \label{HamiUniversal}
\end{equation}
where the field labels refer to two mutually orthogonal polarization unit
vectors for each mode, $\Psi $ and $\Psi ^{\perp },$ corresponding
respectively to the state vectors: $\left| \Psi \right\rangle $ and $\left|
\Psi ^{\perp }\right\rangle .$ This Hamiltonian generates a unitary
transformation $\hat{U}=e^{-i\frac{\hat{H}t}{\hbar }\text{ }}$acting on the
input multimode photon state.

Let's consider first the injection of a polarization encoded {\it qubit }in
the QIOPA\ system:
\begin{equation}
\left| \Psi \right\rangle _{IN}\equiv \alpha \left| \Psi \right\rangle
_{IN}^{H}+\beta \left| \Psi \right\rangle _{IN}^{V}
\end{equation}
with $\left| \alpha \right| ^{2}+\left| \beta \right| ^{2}=1$, defined in
the $2\times 2$-dimensional Hilbert space of polarizations $(\overrightarrow{%
\pi })$. We analyze this configuration by accounting for the $2$ interacting
optical modes ${\bf k}_{1}$ and ${\bf k}_{2}$, i.e. in terms of the basis
vectors: $\left| \Psi \right\rangle _{IN}^{H}=\left| 1\right\rangle
_{1H}\left| 0\right\rangle _{1V}\left| 0\right\rangle _{2H}\left|
0\right\rangle _{2V}\equiv \left| 1,0,0,0\right\rangle $, $\left| \Psi
\right\rangle _{IN}^{V}=\left| 0,1,0,0\right\rangle $. In virtue of the
general \ {\it information preserving }property of any nonlinear (NL)
transformation of parametric type, the output state is again expressed by a
''{\it multiparticle qubit}'':

\begin{equation}
\left| \Psi \right\rangle \equiv \hat{U}\left| \Psi \right\rangle _{IN}=\hat{%
U}\left( \alpha \left| \Psi \right\rangle _{IN}^{H}+\beta \left| \Psi
\right\rangle _{IN}^{V}\right) =\alpha \left| \Psi \right\rangle ^{H}+\beta
\left| \Psi \right\rangle ^{V}  \label{Outputwavefunction}
\end{equation}
Since $\left| \Psi \right\rangle _{IN}$ is a {\it pure} state and $\hat{U}$
is {\it unitary}, $\left| \Psi \right\rangle $ is also a pure state.
Furthermore and most important, it consists of a quantum superposition of
two orthonormal, multiparticle states $\left| \Psi \right\rangle ^{H}$ and $%
\left| \Psi \right\rangle ^{V}$. This is indeed the dominant property
underlying the celebrated {\it Schroedinger Cat }concept{\it \ }\cite
{DeMa98,Schroe35}:

\begin{eqnarray}
\left| \Psi \right\rangle ^{H} &\equiv &\gamma \sum\limits_{i,\ j=0}^{\infty
}(-\Gamma )^{i+j}(-1)^{j}\sqrt{i+1}\left| i+1,j,j,i\right\rangle \\
\left| \Psi \right\rangle ^{V} &\equiv &\gamma \sum\limits_{i,\ j=0}^{\infty
}(-\Gamma )^{i+j}(-1)^{j}\sqrt{j+1}\left| i,j+1,j,i\right\rangle
\end{eqnarray}
where $C\equiv \cosh g$, $\gamma \equiv C^{-3}$, $\Gamma \equiv
\tanh g$, and $g\equiv \kappa \overline{t}_{int}$ expresses the NL
gain of the parametric process, $\overline{t}_{int}$ being the
interaction time.

In order to implement the {\it pair extraction technique }described above in
Section 1, we may now explicitly derive the theoretical expression of the ''%
{\it pair extracted}'' density matrix in regime of photon losses on all
QIOPA spatial and polarization output modes. According to a standard
procedure, the losses are simulated by the insertion of dummy ideal beam
splitters (BS)\ on the optical modes, followed by ideal detectors \cite{Loud}%
.\ Let us first express the above QIOPA\ output wavefunction in
absence of losses, in terms of vacuum  states:

\begin{equation}
\begin{aligned}
& \left| \Psi \right\rangle =\gamma \sum\limits_{i,\ j=0}^{\infty
}(-\Gamma
)^{i+j}\frac{(-1)^{j}}{j!i!}\\&\left( \hat{a}_{1H}^{\dagger (i+1)}\hat{a}%
_{1V}^{\dagger j}\hat{a}_{2H}^{\dagger j}\hat{a}_{2V}^{\dagger i}-\hat{a}%
_{1H}^{\dagger i}\hat{a}_{1V}^{\dagger (j+1)}\hat{a}_{2H}^{\dagger j}\hat{a}%
_{2V}^{\dagger i}\right) \left| 0,0,0,0\right\rangle
\end{aligned}
\end{equation}

\begin{figure}[t]
\includegraphics[scale=.4]{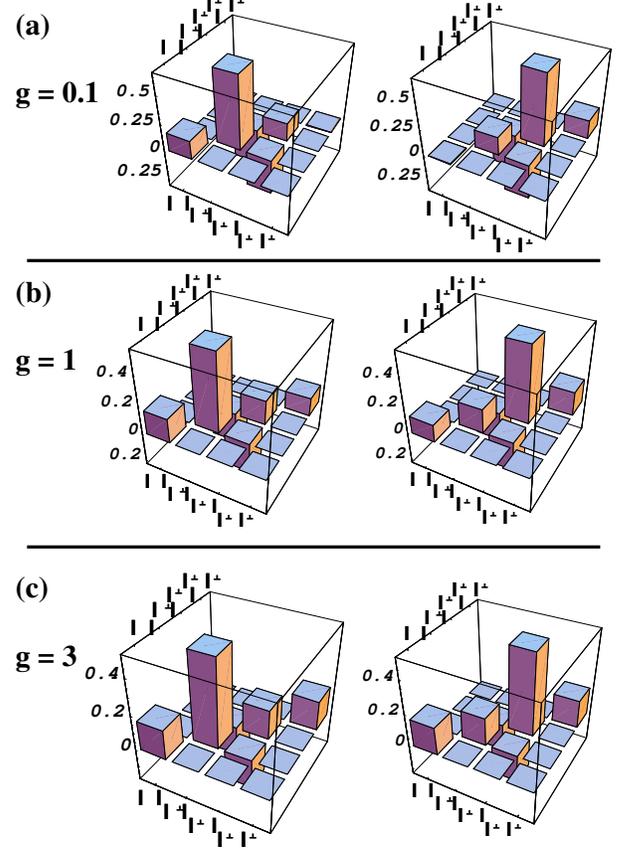} \caption{((Color on line) Theoretical density matrix of the reduced
two-photon output state over the output modes ${\bf k}_{i}$
$\{i=1,2\}$ conditioned by the injection of a generic input
polarization qubit $\left|
\phi \right\rangle $\ (left column) and $\left| \phi ^{\perp }\right\rangle $%
\ (right column) on the mode ${\bf k}_{1}$. (a) corresponds to
$g=0.1$, (b) to $g=1$, (c) to $g=3$. The imaginary parts with all
matrix elements equal to 0 are not reported.} \label{fig2}
\end{figure}

After insertion of the BS's , the lossless QIOPA output field operators $%
\left\{ \hat{a}_{ij}^{\dagger }\right\} $are transformed into the $\left\{
\widehat{a}_{ij-OUT}^{\dagger }\right\} $acting again on the output lossy
channels, by the unitary BS\ map:
\begin{equation}
\left(
\begin{array}{c}
\widehat{a}_{ij-OUT}^{\dagger }(t) \\
\widehat{b}_{ij-OUT}^{\dagger }(t)
\end{array}
\right) =\left(
\begin{array}{cc}
\sqrt{\eta } & \sqrt{1-\eta} \\
\sqrt{1-\eta} & \sqrt{\eta }
\end{array}
\right) \left(
\begin{array}{c}
\hat{a}_{ij}^{\dagger }(t) \\
\widehat{b}_{ij}^{\dagger }(t)
\end{array}
\right)  \label{beamsplitter}
\end{equation}
\newline
Here the operators $\left\{ \widehat{b}_{ij}^{\dagger }\right\} $and $%
\left\{ \widehat{b}_{ij-OUT}^{\dagger }\right\} $ correspond to the input
and output side, i.e.''reflected modes'' of the dummy BS's that do not
coincide with the QIOPA\ states. Precisely, the set $\left\{ \widehat{b}%
_{ij}^{\dagger }\right\} $and $\left\{ \widehat{b}_{ij-OUT}^{\dagger
}\right\} $act correspondingly on a set of side vacuum states and on the set
of ''reflected modes'' $\{\widetilde{{\bf k}}_{i}^{OUT}\}$. We assume
further that all BS's are equal and characterized by a common spatially and
polarization independent transmittivity parameter $\eta $. \newline
The lossy effect of the BS's maps $\left| \Psi \right\rangle $ into an
output state $\left| \Psi \right\rangle _{OUT}^{BS}$ which involves the four
output ''transmitted modes'' and the four output ''reflected modes'' of the
two BS's.\ Precisely, $\left| \Psi \right\rangle _{OUT}^{BS}$ is expressed
in terms of Fock states: $\left| n_{1H},n_{1V},n_{2H},n_{2V}\right\rangle
_{a}\otimes \left| n_{1H},n_{1V},n_{2H},n_{2V}\right\rangle _{b}$ where the
two terms in the tensor product represent respectively the output
transmitted BS's modes ($\widehat{a}-$modes) and the output reflected modes (%
$\widehat{b}-$modes). The output state of the overall $\left\{
QIOPA+BS' s\right\} $ system, is found to reproduce the quantum
superposition behavior:
\begin{widetext}
\begin{equation}
\left| \Psi \right\rangle _{OUT}^{BS}=\gamma \sum\limits_{i,j=0}^{\infty }%
\frac{(-\Gamma )^{i+j}(-1)^{j}}{i!j!}\left( -\jmath \sqrt{1-\eta ^{2}}%
\right) ^{2\ast (i+j)+1}\left( \alpha \left| \Phi _{i,j}^{H}\right\rangle
+\beta \left| \Phi _{i,j}^{V}\right\rangle \right)
\end{equation}
where
\begin{eqnarray}
\left| \Phi _{i,j}^{H}\right\rangle
&=&\sum_{l_{1}=0}^{i+1}\sum_{l_{2}=0}^{j}\sum_{l_{3}=0}^{j}%
\sum_{l_{4}=0}^{i}\left( \frac{\eta }{-\jmath \sqrt{1-\eta
^{2}}}\right)
^{(l_{1}+l_{2}+l_{3}+l_{4})}j!\sqrt{%
{i+1 \choose l_{1}}%
{j \choose l_{2}}%
{j \choose l_{3}}%
{i \choose l_{4}}%
(i+1)!i!} \\
&&\times \left| l_{1},l_{2},l_{3},l_{4}\right\rangle _{a}\left|
i+1-l_{1},j-l_{2},j-l_{3},i-l_{4}\right\rangle _{b}  \nonumber \\
\left| \Phi _{i,j}^{V}\right\rangle
&=&\sum_{l_{1}=0}^{i}\sum_{l_{2}=0}^{j+1}\sum_{l_{3}=0}^{j}%
\sum_{l_{4}=0}^{i}\left( \frac{\eta }{-\jmath \sqrt{1-\eta
^{2}}}\right)
^{(l_{1}+l_{2}+l_{3}+l_{4})}i!\sqrt{%
{i \choose l_{1}}%
{j+1 \choose l_{2}}%
{j \choose l_{3}}%
{i \choose l_{4}}%
(j+1)!j!} \\
&&\times \left| l_{1},l_{2},l_{3},l_{4}\right\rangle _{a}\left|
i-l_{1},j+1-l_{2},j-l_{3},i-l_{4}\right\rangle _{b}  \nonumber
\end{eqnarray}
\end{widetext}
 with $\jmath^{2}=-1$. All undetected BS ''reflected modes'' must then be discarded,
i.e. traced
out in the above expressions leading to the {\it reduced }density matrix $%
\rho ^{\prime }=Tr_{b}\left[ \left| \Psi \right\rangle
_{OUT}^{BS}\,_{OUT}^{BS}\left\langle \Psi \right| \right] $ over the
transmitted modes $\{{\bf k}_{i}^{OUT}\}\ $that can be expressed as follows:
\begin{equation}
\rho ^{\prime }=\sum_{k_{1}=0}^{\infty }\sum_{k_{2}=0}^{\infty
}\sum_{k_{3}=0}^{\infty }\sum_{k_{4}=0}^{\infty }\left\langle
k_{1},k_{2},k_{3},k_{4}\right| _{b}\left| \Psi \right\rangle
_{OUT}^{BS}\left\langle \Psi \right| \left|
k_{1},k_{2},k_{3},k_{4}\right\rangle _{b}
\end{equation}

Up to now we have considered arbitrary, polarization-symmetric
channel losses. In the following we make the additional assumption
of high losses, which will greatly simplify the calculations. The
explicit expression of $\ \rho ^{\prime }$ may now be easily
obtained in the high loss regime, i.e. identified by the relation
$\eta \overline{n}<<1$, $\eta \overline{n}$ being the average number
of photons transmitted by the BS per mode.\ In virtue of this
condition we may drop the terms of the sums proportional to $\eta
^{n}$ for $n>2$. This corresponds to consider only matrix elements
of a representation in which any Fock basis state correspond to a
photon occupation number $n\leq 2$. As final step we assume to
detect one photon on each mode ${\bf k}_{i}^{OUT}$ by a standard
2-photon coincidence technique over the QIOPA\ output modes, ${\bf
k}_{1}$and ${\bf k}_{2}$. By this technique we only investigate
output states involving 1 photon emitted over these two modes. This
output condition is expressed by a matrix representation of $\ \rho
^{\prime }\;$involving the basis states:$\left\{ \left|
1,0,1,0\right\rangle ,\left| 1,0,0,1\right\rangle ,\left|
0,1,1,0\right\rangle ,\left| 0,1,0,1\right\rangle \right\} $,
corresponding to the basis: $\left\{ \left| H\right\rangle
_{1}\left| H\right\rangle _{2},\left| H\right\rangle _{1}\left|
V\right\rangle _{2},\left| V\right\rangle _{1}\left| H\right\rangle
_{2},\left| V\right\rangle _{1}\left| V\right\rangle _{2}\right\} $.
The explicit expression of the detected ''pair extracted'' density
matrix is finally given by the expression:
\begin{widetext}
\begin{equation}
\rho ^{\prime }=\frac{1}{3d}\left(
\begin{array}{cccc}
\left| \alpha \right| ^{2}d+\left| \beta \right| ^{2}2t^{2} & \beta \alpha
^{\ast }d & -\beta \alpha ^{\ast }2 & 0 \\
\alpha \beta ^{\ast }d & \left| \alpha \right| ^{2}2(1+d)+\left| \beta
\right| ^{2}d & -\left| \alpha \right| ^{2}2-\left| \beta \right| ^{2}2 &
\beta \alpha ^{\ast }2 \\
-\alpha \beta ^{\ast }2 & -\left| \alpha \right| ^{2}2-\left| \beta \right|
^{2}2 & \left| \alpha \right| ^{2}d+\left| \beta \right| ^{2}2(1+d) & -\beta
\alpha ^{\ast }d \\
0 & \alpha \beta ^{\ast }2 & -\alpha \beta ^{\ast }d & \left| \alpha \right|
^{2}2t^{2}+\left| \beta \right| ^{2}d
\end{array}
\right)  \label{RhoOut}
\end{equation}
\end{widetext}
with $d=(1+t^{2})$ and $t=\Gamma (1-\eta ^{2})$. Let
us consider the particular asymptotic case $\eta \rightarrow 0$,
i.e., $t\simeq \tanh g$. Figure 2 refers to different injection
states and to different values of the interaction parameter $g$.
There the tomographic patterns are reproduced identically for any
couple of different orthogonal input states $\left\{ \left| \phi
\right\rangle ,\left| \phi ^{\bot }\right\rangle \right\} $ as a
consequence of the universality of this process,
Eq.(\ref{HamiUniversal}). Furthermore, the structure of the $4\times
4$ matrices shows the relevant quantum features of the output state.
For instance, the highest peak on the diagonals expressing the
quantum superposition of the input state shifts from the position
$\left| \phi \phi ^{\bot }\right\rangle \left\langle \phi \phi
^{\bot }\right| $ to $\left| \phi ^{\bot }\phi \right\rangle
\left\langle \phi ^{\bot }\phi \right| $ in correspondence with the
OPA excitation by any set of generic orthogonal injection states
$\left\{ \left| \phi \right\rangle ,\left| \phi ^{\bot
}\right\rangle \right\} $.

Let us estimate the entanglement of the {\it pair extracted} output state (%
\ref{RhoOut}). Owing to the tested universality of the amplification process
we restrict the present analysis to the input qubit $\ \left| \Psi
\right\rangle _{IN}\equiv \left| H\right\rangle $, i.e. $\alpha =1,$ $\beta
=0$. In this case the two-photon density matrix reads:
\begin{equation}
\rho ^{\prime }=\frac{1}{3(1+t^{2})}\left(
\begin{array}{cccc}
\frac{1}{2}(1+t^{2}) & 0 & 0 & 0 \\
0 & (2+t^{2}) & -1 & 0 \\
0 & -1 & \frac{1}{2}(1+t^{2}) & 0 \\
0 & 0 & 0 & t^{2}
\end{array}
\right)  \label{RhoH}
\end{equation}
The entanglement measurement, the ''concurrence'' is $C(\rho ^{\prime })=%
\frac{2}{3(1+t^{2})}\left( 1-\frac{t}{2}\sqrt{1+t^{2}}\right) $\cite
{Wootters98}. We found $C(\rho ^{\prime })>0$ for any value of $g$ and $%
C(\rho ^{\prime })\rightarrow 0$ for $g\rightarrow \infty .$ This
result complies with the one obtained in the case of the
$1\longrightarrow M$ universal quantum cloning by \cite{Brus03}.
There it was shown that, in this process closely parallel to the
present configuration, the concurrence between a clone and an
ancilla is always different from zero for any value of $M$ and
vanishes in the limit $M\rightarrow \infty $.

We now analyze the output state (\ref{RhoH}) over each one of the output
modes, by further tracing $\rho ^{\prime }$. For the mode ${\bf k}_{1}$ we
obtain:
\begin{equation}
\rho _{k1}^{\prime }=\frac{1}{6(1+t^{2})}\left(
\begin{array}{cc}
5+3t^{2} & 0 \\
0 & 1+3t^{2}
\end{array}
\right)
\end{equation}
The fidelity between the output state and the input is: $F(\left|
H\right\rangle ,\rho _{k1}^{\prime })=\left\langle H\right| \rho
_{k1}^{\prime }\left| H\right\rangle =\frac{5+3t^{2}}{6(1+t^{2})}$. In the
limit $g\rightarrow \infty $ we get $F=\frac{2}{3}$while the limit $%
g\rightarrow 0$ leads to $F=\frac{5}{6}$. In the present investigation,
dealing with the generation of a pair of photons, we retrieved exactly the
result obtained for the $1\rightarrow 2$ optimal cloning process. The
average number of photon over the mode ${\bf k}_{1}$ is equal to $3\overline{%
n}+1$ with $\overline{n}=\sinh ^{2}g$. Hereafter the mode ${\bf k}_{1}$ will
also be referred to as the {\it cloning mode}.

The single photon polarization state over mode ${\bf k}_{2}$ is described by
the density matrix for any $g$ value:
\begin{equation}
\rho _{k2}^{\prime }=\left(
\begin{array}{cc}
1/3 & 0 \\
0 & 2/3
\end{array}
\right)
\end{equation}
The fidelity between the output state and the orthogonal of the input one $%
F(\left| V\right\rangle ,\rho _{k2}^{\prime })=2/3$. Hereafter the mode $%
{\bf k}_{2}$ will be referred to as the {\it anti-cloning mode}.

\section{Schroedinger Cat State: persistence of entanglement}

\begin{figure}[t]
\includegraphics[scale=.3]{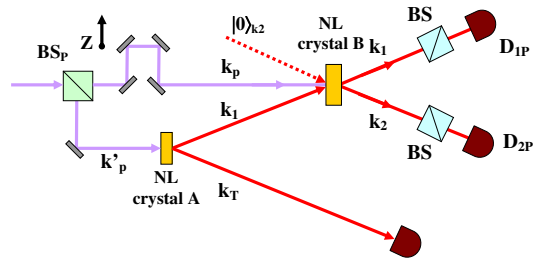} \caption{(Color on line) Schematic diagram of the {\it quantum injected}
optical parametric amplifier (QIOPA). The pump beam ${\bf
k}_{P}^{\prime }$ excites the NL crystal A which generates pairs of
entangled photons over the modes ${\bf k}_{1}$ and ${\bf k}_{T}$.
The photon over the mode ${\bf k}_{T}$ provides the trigger of the
overall experiment while the twin photon over
the mode ${\bf k}_{1}$ excites the NL crystal B together with the pump beam $%
{\bf k}_{P}$ \cite{DeMa98}. The losses are simulated by the
insertion of a beamsplitter $(BS)$ over each propagation mode ${\bf
k}_{i}$.} \label{fig2}
\end{figure}

Let's now consider a different, more complex configuration in which the
injected photon belongs to a polarization entangled pair: see Fig.3. The
previous theoretical method will be here adopted to investigate the
properties of the overall output field. Let us now go into details. The
initial pump beam is split by an unbalanced beam-splitter ($BS_{P}$) into
two beams. A low intensity beam ${\bf k}_{P}^{\prime }$ excites the NL
crystal A which generates pair of entangled photons over the modes ${\bf k}%
_{1}$ and ${\bf k}_{T}$:
\begin{equation}
\left| \Phi ^{-}\right\rangle _{kT,k1}=2^{-%
{\frac12}%
}\left( \left| H\right\rangle _{kT}\left| H\right\rangle _{k1}-\left|
V\right\rangle _{kT}\left| V\right\rangle _{k1}\right)
\end{equation}
where subscripts ${\bf k}_{T}$, ${\bf k}_{1}$ refer to trigger and injection
photons, propagating along ${\bf k}_{T}$ and ${\bf k}_{1}$ modes,
respectively. The pump power is sufficiently low to avoid the simultaneous
generation of more than one pair of photons. The photon over the mode ${\bf k%
}_{T}$ provides the trigger of the overall experiment while the twin photon
over the mode ${\bf k}_{1}$ excites the NL crystal B together with the pump
beam ${\bf k}_{P}$ \cite{DeMa98}.

The overall dynamic of the process is described by the unitary operator $%
\widehat{V}=\widehat{I}_{kT}\otimes \widehat{U}_{k1}$ acting on the initial
state $\left| \Phi ^{-}\right\rangle _{Tk1}\otimes \left| 0\right\rangle
_{k2}$, where $\widehat{U}_{k1}$ is the time evolution operator acting on
the injection state and $\widehat{I}_{kT}$\ is the unit matrix acting on
trigger state. Using the results of the previous section we obtain

\begin{eqnarray}
\left| \Sigma \right\rangle &=&\widehat{V}\left| \Phi ^{-}\right\rangle
_{kT,k1}=\\
&&2^{-%
{\frac12}%
}\gamma \{\left| H\right\rangle _{kT}\otimes \sum\limits_{i,j=0}^{\infty
}(-\Gamma )^{i+j}(-1)^{j}\sqrt{i+1}\left| i+1,j,j,i\right\rangle
\label{Wavefunction} \\
&&-\left| V\right\rangle _{kT}\otimes \sum\limits_{i,j=0}^{\infty }(-\Gamma
)^{i+j}(-1)^{j}\sqrt{j+1}\left| i,j+1,j,i\right\rangle \}
\end{eqnarray}
This multiparticle quantum state exhibits an entangled structure connecting
the microscopic property of the system, i.e. the single particle ''trigger''
acting on the mode ${\bf k}_{T}$, and the macroscopic quantum superposition.
This indeed corresponds to the original definition of ''{\it %
Schroedinger Cat State''} \cite{Schroe35,Schleich01}.

\begin{figure}[t]
\includegraphics[scale=.4]{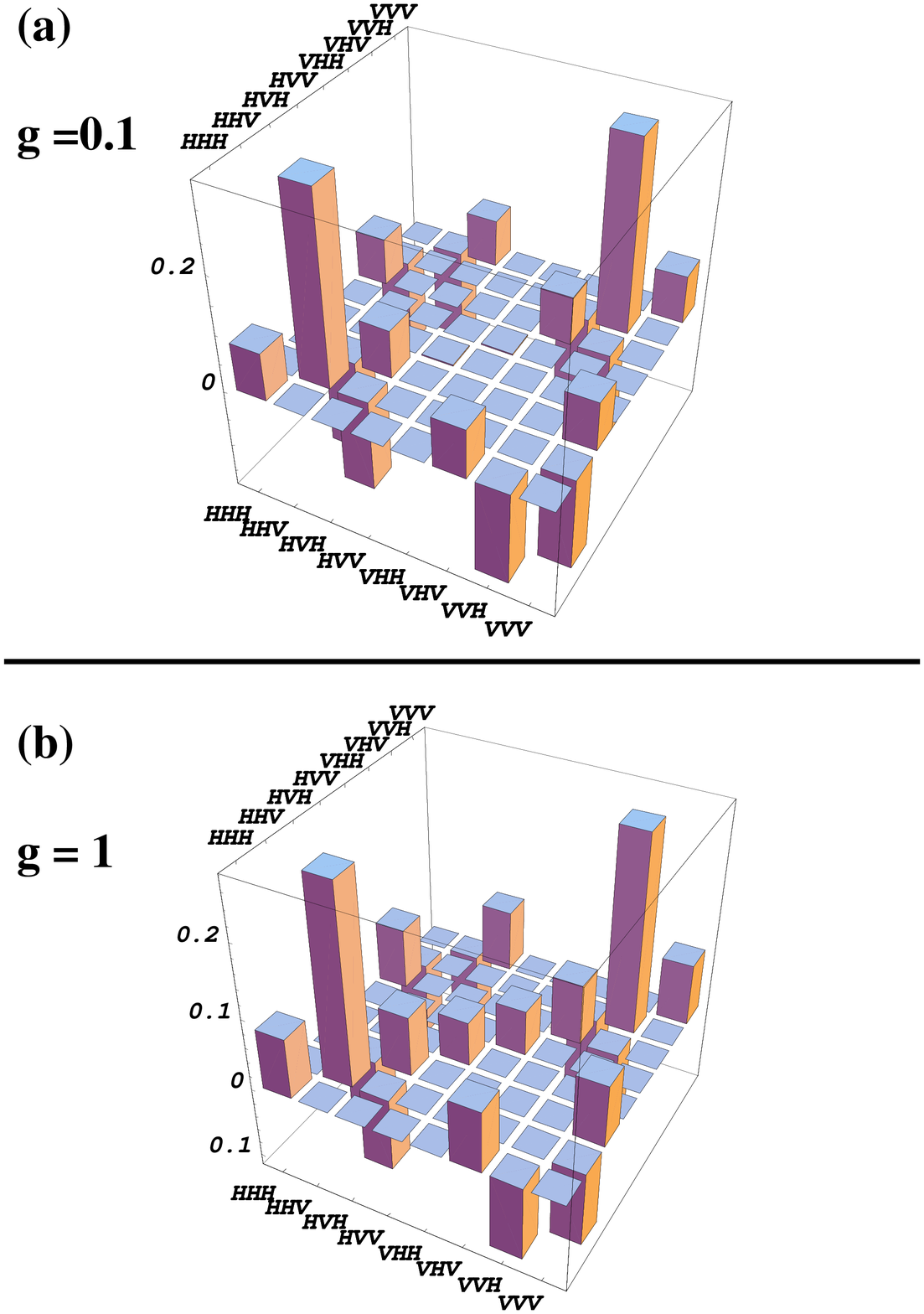} \caption{(Color on line) Theoretical density matrix of the reduced
three-photon output state over the modes ${\bf k}_{T}$, ${\bf k}_{i}$ $%
\{i=1,2\}$ (a) corresponds to $g=0.1$, (b) to $g=1$. The imaginary
parts with all matrix elements equal to 0 are not reported.}
\label{fig2}
\end{figure}

In the following we apply the\ {\it pair extraction} method to
analyze the entanglement properties of the overall wavefunction
(\ref{Wavefunction}). High losses are introduced over the modes
${\bf k}_{i}$ $\{i=1.2\}$. Detection of photons over the three modes
${\bf k}_{i}$, ${\bf k}_{T}$ enables to reconstruct the density
matrix through a tomographic technique. The theoretical model shows
the presence of entanglement between the trigger photon and each one
of the amplified photons. Since the entanglement is not ascribable
to the subsequent amplification process which is acting locally on
the ${\bf k}_{1}$ arm of the pair, one is forced to conclude that
the original trigger-injection entanglement has survived the QIOPA\
amplification.

Under the {\it pair extraction} approximation on ${\bf k}_{i}$ $\{i=1.2\}$,
i.e. after applying high losses on the output modes before detection, the
expression of the overall normalized 3 qubit {\it pair extracted} density
matrix is easily found:
\begin{widetext}
\begin{equation}
\rho ^{\prime \prime }=\left(
\begin{array}{cccccccc}
\frac{1}{12} & 0 & 0 & 0 & 0 & \frac{1}{12} & -\frac{1}{6+6t^{2}} & 0 \\
0 & \frac{2+t^{2}}{6+6t^{2}} & -\frac{1}{6+6t^{2}} & 0 & 0 & 0 & 0 & -\frac{1%
}{6+6t^{2}} \\
0 & -\frac{1}{6+6t^{2}} & \frac{1}{12} & 0 & 0 & 0 & 0 & \frac{1}{12} \\
0 & 0 & 0 & \frac{t^{2}}{6+6t^{2}} & 0 & 0 & 0 & 0 \\
0 & 0 & 0 & 0 & \frac{t^{2}}{6+6t^{2}} & 0 & 0 & 0 \\
\frac{1}{12} & 0 & 0 & 0 & 0 & \frac{1}{12} & -\frac{1}{6+6t^{2}} & 0 \\
-\frac{1}{6+6t^{2}} & 0 & 0 & 0 & 0 & -\frac{1}{6+6t^{2}} & \frac{2+t^{2}}{%
6+6t^{2}} & 0 \\
0 & -\frac{1}{6+6t^{2}} & \frac{1}{12} & 0 & 0 & 0 & 0 & \frac{1}{12}
\end{array}
\right)  \label{qubit3}
\end{equation}
\end{widetext}
Figure 4 reports the various density matrices for
different values of interaction parameter $g$.

In order to understand at a deeper level the multiparticle quantum cloning
process, we may consider the $4\times 4$ dimensional matrices obtained by
tracing $\rho ^{\prime \prime }$ over the remaining state of the manifold $%
\left\{ {\bf k}_{i},{\bf k}_{T}\right\} .$

Let us start from the system consisting of the ${\bf k}_{i}$ $\{i=1.2\}$
modes, i.e. by the cloning-anticloning systems. The reduced density matrix $%
\rho _{k1,k2}^{\prime \prime }=Tr_{kT}\left( \rho ^{\prime \prime }\right) $
of the extracted 2-photons is found:
\begin{equation}
\rho _{k1,k2}^{\prime \prime }=\left(
\begin{array}{cccc}
\frac{1-p}{4} & 0 & 0 & 0 \\
0 & \frac{1+p}{4} & -\frac{p}{2} & 0 \\
0 & -\frac{p}{2} & \frac{1+p}{4} & 0 \\
0 & 0 & 0 & \frac{1-p}{4}
\end{array}
\right)  \label{cloninganticloning}
\end{equation}
with $p=\frac{2}{3}\frac{1}{1+t^{2}}$. We note that the above density matrix
has the form of a Werner state (WS) $\rho _{W}=p\left| \Psi
_{-}\right\rangle \left\langle \Psi _{-}\right| +\frac{1-p}{4}I$, i.e. a
mixture of the maximally entangled state$\ \left| \Psi _{-}\right\rangle
_{1,2}=2^{-1/2}(\left| H\right\rangle _{1}\left| V\right\rangle _{2}-\left|
V\right\rangle _{1}\left| H\right\rangle _{2})$ with probability $p$ and of
the\ fully chaotic state $I/4$, being $I$ the identity operator on the
overall Hilbert space. Note that for any finite value of the NL value $g$,
the singlet probability is $p>\frac{1}{3}$ reaching asymptotically the value
$p=\frac{1}{3}$ for $g\rightarrow \infty $. Since the condition $p>\frac{1}{3%
}$ is a necessary and sufficient one for state non-separability of any WS
\cite{Barb04}, the entanglement condition affecting the {\it pair extracted}
$\rho _{k1,k2}^{\prime \prime }$ , and then of its multiparticle counterpart
$\rho _{k1,k2}$, {\it does persist} for any finite value of $g$. All this
can be compared with a similar result obtained in regime of spontaneous
parametric down-conversion. i.e., with no injection of a single photon
state. In the SPDC case it was found: $\overline{p}=\frac{1}{1+2t^{2}}$ \cite
{Cami06}. The concurrence of the state (\ref{cloninganticloning}) is $%
C_{k1,k2}=2\max \left( \frac{3p-1}{4},0\right) =\frac{1}{2}\left( \frac{%
1-t^{2}}{1+t^{2}}\right) =\frac{1}{2}(\sinh ^{2}g+\cos ^{2}g)^{-1}$. In case
of high gain value we find $C_{k1,k2}\simeq \frac{1}{\overline{n}}$ and $%
\tau _{kT,k1}=\frac{1}{C_{T,k1}^{2}}=\frac{1}{\overline{n}^{2}}$. The
average number of output qubits for each mode is equal to $\simeq 3\overline{%
n}.$

A similar analysis can be carried out for the correlation affecting the
modes ${\bf k}_{T}$ and ${\bf k}_{1}$, as shown by the structure of the pair
extracted $\rho _{kT,k1}^{\prime \prime }\equiv Tr_{k2}\left( \rho ^{\prime
\prime }\right) $:
\begin{equation}
\rho _{kT,k1}^{"}=\left(
\begin{array}{cccc}
\frac{1+q}{4} & 0 & 0 & -\frac{q}{2} \\
0 & \frac{1-q}{4} & 0 & 0 \\
0 & 0 & \frac{1-q}{4} & 0 \\
-\frac{q}{2} & 0 & 0 & \frac{1+q}{4}
\end{array}
\right)  \label{triggercloning}
\end{equation}
with $q=\frac{2}{3}\frac{1}{1+t^{2}}$. Note that once again the density
matrix bears the WS\ structure $\rho =q\left| \Phi _{-}\right\rangle
\left\langle \Phi _{-}\right| +\frac{1-q}{4}I$, i.e. is a mixture with
probability $q$ of the maximally entangled state$\ \left| \Phi
_{-}\right\rangle _{1,2}=2^{-1/2}(\left| H\right\rangle _{1}\left|
H\right\rangle _{2}-\left| V\right\rangle _{1}\left| V\right\rangle _{2})$
and of the\ maximally chaotic state $I/4$. Once again the entanglement
condition $q>\frac{1}{3}$ is met for any finite value of $g$ and $%
q\rightarrow \frac{1}{3}$ for $g\rightarrow \infty $. Note also that in the
limit of small $g\approx 0$ is: $q\approx \frac{2}{3}$ (the present approach
considers at least the generation of a pair of photons). The concurrence of
the state of Eq.(\ref{triggercloning}) is found: $C_{kT,k1}=\frac{1}{2}%
(\sinh ^{2}g+\cosh ^{2}g)^{-1}$. In case of large gain value is found: $%
C_{kT,k1}\simeq \frac{1}{\overline{n}}$ and $\tau _{kT,k1}=C_{kT,k1}^{2}=%
\frac{1}{\overline{n}^{2}}$ . The average number of output clones is equal
to $3\overline{n}+1\simeq 3\overline{n}$.

\begin{figure}[t]
\includegraphics[scale=.4]{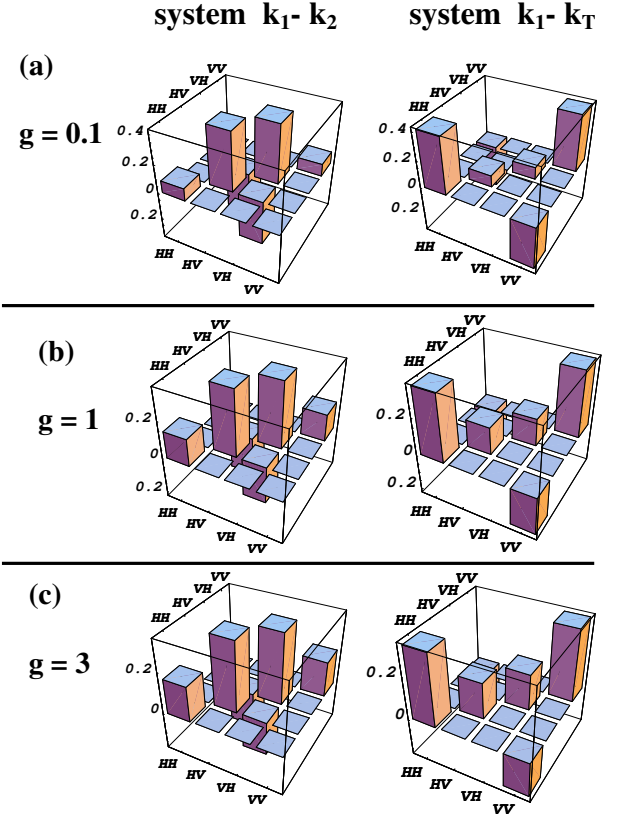} \caption{(
Figure 5. (Color on line) Theoretical density matrix of the reduced
two-photon output state over the output modes ${\bf k}_{1},$ ${\bf
k}_{2}$ (left column) and ${\bf k}_{1},$ ${\bf k}_{T}$\ (right
column). (a) corresponds to $g=0.1$, (b) to $g=1$, (c) to $g=3$. The
imaginary parts with all matrix elements equal to 0 are not
reported.} \label{fig2}
\end{figure}

Consider at last the pair extracted reduced density matrix involving the $%
{\bf k}_{2}$ and ${\bf k}_{T}$ modes: $\rho _{kT,k2}^{\prime \prime }\equiv
Tr_{k1}\left( \rho ^{\prime \prime }\right) $. Its expression is found to be
gain independent
\begin{equation}
\rho _{kT,k2}^{\prime \prime }=\left(
\begin{array}{cccc}
\frac{1}{6} & 0 & 0 & \frac{1}{6} \\
0 & \frac{1}{3} & 0 & 0 \\
0 & 0 & \frac{1}{3} & 0 \\
\frac{1}{6} & 0 & 0 & \frac{1}{6}
\end{array}
\right)  \label{trigger-ac}
\end{equation}
It also exhibits a WS structure, $\rho _{kT,k2}^{\prime \prime }=l\left|
\Phi _{+}\right\rangle \left\langle \Phi _{+}\right| +\frac{1-l}{4}I$ with $%
l=\frac{1}{3}$ and $\left| \Phi _{+}\right\rangle _{1,2}=2^{-1/2}(\left|
H\right\rangle _{1}\left| H\right\rangle _{2}+\left| V\right\rangle
_{1}\left| V\right\rangle _{2})$. As a consequence, the above state is {\it %
separable, }a result{\it \ }in agreement with the theory of the optimal
quantum machines \cite{Scia04}. Indeed the QIOPA has been shown to implement
over the mode ${\bf k}_{2}$ the universal optimal flipping machine \cite
{Pell03}. The completely positive (CP) map which implements the UNOT gate
has the following Kraus representation ${\cal E}_{UNOT}(\rho )=\frac{1}{3}%
\left( \sigma _{X}\rho \sigma _{X}+\sigma _{Y}\rho \sigma _{Y}+\sigma
_{Z}\rho \sigma _{Z}\right) $. The state $\rho _{kT,k2}^{\prime \prime }$
can be obtained applying to the entangled state $\left| \Phi
^{-}\right\rangle _{kT,k1}$ the identity over the mode ${\bf k}_{T}$ and the
CP map ${\cal E}_{UNOT}$ over the mode ${\bf k}_{1}$:

\begin{equation}
\begin{aligned}
&I_{kT}\otimes {\cal E}_{UNOT-k1}\left( \left| \Phi
^{-}\right\rangle _{kT,k1}\left\langle \Phi ^{-}\right|
_{kT,k1}\right) \\&=\frac{1}{3}\left[
\begin{array}{c}
\left| \Phi ^{+}\right\rangle _{kT,k2}\left\langle \Phi ^{+}\right|
_{kT,k2}+\left| \Psi ^{-}\right\rangle _{kT,k2}\left\langle \Psi ^{-}\right|
_{kT,k2}+ \\
+\left| \Psi ^{+}\right\rangle _{kT,k2}\left\langle \Psi ^{+}\right| _{kT,k2}
\end{array}
\right]
\end{aligned}
\end{equation}
Interestingly enough, the last expression can be shown to be equal to Eq.(%
\ref{trigger-ac}). This confirms the overall validity of the present
approach.

\section{Discussions and conclusions}

In the present paper, theory of quantum injected optical parametric
amplification has been extensively investigated in regime of
high-gain and high losses with particular attention for the
entanglement properties of the output fields. We have exploited the
developed theoretical tool to investigate the properties of
entangled states after a cloning process, demonstrating the
persistence of entanglement for any clone-trigger subsystem.
Connections with Werner state have been established in the physical
process of stimulated emission. In addition, we have shown that the
properties of \ $QIOPA$ output state do comply with the ones
conventionally expected for a Schroedinger Cat system
\cite{Schleich01,Schroe35}. With respect to the persistence of the
coherence of the latter system for increasing ''size'', an
interesting problem could be the investigation on the entanglement
persistence for increasing the value of the gain $g$, i.e. for a
very large number of generated clones. Unfortunately we expect that,
for an increasing number of clones, the small amount of bipartite
entanglement should not be observable due to experimental
imperfections (like walk off effects and other sources of
decoherence). However the density matrices experimentally measured
in Ref.\cite{Eise04,DeMa05b,Cami06} have been found in good
agreement with the theoretical ones. On the other hand, the
technique introduced above turns out to be an useful tool to
investigate single photon features of mesoscopic fields.

An interesting aspect which deserves further investigation is the
effect of losses on the orthogonality condition of the two wave
functions of Eq.5 and 6, in particular how the detection efficiency
influences the distinction of the two initial orthogonal terms. The
two extreme condition can easily been derived. While for vanishing
losses the Hilbert-Schimdt distance of the two interfering state is
found to be
$d(|\Psi\rangle^{H},|\Psi\rangle^{V})=Tr[(|\Psi\rangle^{H}\langle\Psi|-|\Psi\rangle^{V}\langle\Psi|)^2]=2$,
the condition of very low efficiency of detection leads to
$d(\rho'^{H}_{k1},\rho'^{V}_{k1})=2/9$. Finally a new approach to
investigate quantum features of optical field based on the
combination of data obtained with different detection efficiencies
\cite{Ross04} could improve the present results.

The theoretical results obtained above have been carefully tested by the
experiment reported in Ref.\cite{DeMa05b}. The adopted optical scheme was
similar to the diagram shown in Figure 3 but for a more compact ''folded''
configurations by which a single NL crystal slab, excited in both directions
by the UV\ ''pump'' laser beam, realized in sequence the SPDC\ and the
QI-OPA\ operations. The experimental investigation of the multiphoton
superposition and entanglement implied by Eqs \ref{RhoH}, \ref{qubit3} was
carried out by means of quantum state tomography (QST) according to the {\it %
pair extraction} method previously described. The beams associated with the
output modes ${\bf k}_{i}(i$=$1,2)$ were highly attenuated to the
single-photon level by the two low transmittivity BS's. Experimentally the
maximal value of gain obtained has been found $g_{\exp }=\left( 1.19\pm
0.05\right) $, while the detection quantum efficiencies: read $\eta _{1}$=$%
(4.9\pm 0.2)\%$; $\eta _{2}$=$(4.2\pm 0.2)\%$. By the previous values we
found the condition $\eta \overline{n}\simeq 0.1$. The QST analysis of the
{\it reduced }output{\it \ }state $\rho _{k1,k2}^{\prime \prime }$
determined by the set of input $\left| \Psi \right\rangle _{in}$: $\{\left|
H\right\rangle $, $\left| V\right\rangle $, $\left| \pm \right\rangle \}$.
The good agreement between theory $\rho _{k1,k2}^{\prime \prime }$ and
experiment $\rho _{k1,k2}^{\prime \prime \exp }$ is expressed by the
measured average Uhlmann ''fidelity'': ${\cal F}(\rho _{k1,k2}^{\prime
\prime \exp },\rho _{k1,k2}^{\prime \prime })\equiv \lbrack Tr(\sqrt{\rho
_{k1,k2}^{\prime \prime \exp }}\rho _{k1,k2}^{\prime \prime }\sqrt{\rho
_{k1,k2}^{\prime \prime \exp }})^{%
{\frac12}%
}]^{2}$=$(96.6\pm 1.2)\%$. Finally, the QST\ reconstruction of the density
matrix $\rho ^{\prime \prime \exp }$ has been carried out with a fidelity: $%
{\cal F}(\rho ^{\prime \prime \exp },\rho ^{\prime \prime })$=
$(85.0\pm 1.1)\%$.

Work\ supported by Ministero dell'Istruzione, dell'Universit\`{a} e della
Ricerca (PRIN 2005).

\end{document}